\documentclass{aa}
\usepackage{graphics,amssymb,natbib}
\usepackage{epsfig,times}

\makeatletter
\def\@cite#1#2{(#1\if@tempswa , #2\fi)}
\def\@citex[#1]#2{\if@filesw\immediate\write\@auxout{\string\citation{#2}}\fi
  \def\@citea{}\@cite{\@for\@citeb:=#2\do
    {\@citea\def\@citea{;\penalty\@m\ }\@ifundefined
       {b@\@citeb}{{\bf ?}\@warning
       {Citation `\@citeb' on page \thepage \space undefined}}%
\hbox{\csname b@\@citeb\endcsname}}}{#1}}
\makeatother

\def\xspec{{\sc xspec}}
\def\bb{{\sc bb}}
\def\ktbb{kT_{\rm bb}}
\def\rbb{R_{\rm bb}}

\def\ktsh{kT_{\rm s}^h}

\def\kteh{kT_{\rm e}^h}
\def\ktew{kT_{\rm e}^w}

\def\kts{kT_{\rm s}}
\def\kte{kT_{\rm e}}
\def\dbb{{\sc dbb}}

\def\eqpair{{\sc eqpair}}
\def\comptt{{\sc comptt}}

\def\pl{{\sc pl}}

\def\gaussian{{\sc gaussian}}
\def\covfact{{\Omega/2\pi}}
\def\chiq{$\chi^2$}
\def\cm2{cm$^{-2}$}
\def\s1{s$^{-1}$}

\def\deltam{$\dot{M}$}

\def\wabs{{\sc wabs}}
\def\sax{{\it BeppoSAX}}
\def\gmin{\gamma_{\rm min}}
\def\gmax{\gamma_{\rm max}}
\def\xspec{{\sc xspec}}
\def\eqpair{{\sc eqpair}}
\def\ls{$\ell_{\rm s}$}
\def\lh{$\ell_{\rm h}$}
\def\lth{$\ell_{\rm th}$}
\def\lnth{$\ell_{\rm nth}$}
\def\gierlinski{Gierli\'nski}

\begin{document}

\title{The transient hard X-ray tail of GX 17+2: new \sax\ results}

\author{R. Farinelli\inst{1}, F. Frontera\inst{1,2}, A. A. Zdziarski\inst{3}, L. Stella\inst{4}, S. N. Zhang\inst{5,6}, M. van der Klis\inst{7}, N. Masetti\inst{2}, and L. Amati\inst{2}}

\offprints{R. Farinelli \\\email{farinelli@fe.infn.it}}	  
\titlerunning{The transient hard X-ray tail of GX 17+2}
\authorrunning{R. Farinelli et al.}

\institute{$^1$Dipartimento di Fisica, Universit\`a di Ferrara, via Paradiso
12, 44100 Ferrara, Italy\\ 
$^2$Istituto di Astrofisica Spaziale e Fisica Cosmica, Sezione di Bologna, CNR, 
via Gobetti 101, 40129 Bologna, Italy\\
$^3$N. Copernicus Astronomical Center, Bartycka 18, 00-716 Warsaw, Poland\\
$^4$Osservatorio Astronomico di Roma, via Frascati 33, 00040 Monteporzio Catone, Italy\\
$^5$Physics Department, University of Alabama in Huntsville, AL 35899 Huntsville, USA\\
$^6$Space Sciences Laboratory, NASA Marshall Space Flight Center, SD50, AL 35812 Huntsville, USA\\
$^7$Astronomical Institute ``Anton Pannekoek'', University of Amsterdam, Kruislaan 403, 1098 SJ Amsterdam, The Netherlands\\
}

\date{Received 2 July 2004 \date; / Accepted 17 December 2004}

\abstract{We report on results of two \sax\ observations of  the Z 
source \mbox{GX 17+2}. In both cases the source is in the horizontal branch
of the colour-intensity diagram.
 The persistent continuum can be fit  by  two-component
 models consisting of a blackbody plus a Comptonization spectrum.
With one of these models, two solutions for the blackbody temperature of
both the observed and seed photons for Comptonization are equally accepted
by the data. In the first observation, when the source is on the left part of the horizontal branch,
we observe a hard tail extending up to 120 keV, while 
in the second observation,  when the source moves towards right in the same branch,
the tail is no longer detected. 
The hard ($\ga 30$ keV) \mbox{X-ray} emission can be modeled either by a simple power-law with
photon index $\Gamma \sim$ 2.7, or assuming Comptonization of $\sim$ 1 keV soft photons 
off a hybrid thermal plus non-thermal electron plasma.
The spectral index of the non-thermal injected electrons is $p \sim 1.7$.
The observation of hard \mbox{X-ray} emission only in the left part of the horizontal branch could be indicative of the presence of a threshold in  the accretion rate above which the hard tail disappears. 
 An emission line at 6.7 keV with  equivalent width  $\sim 30$ eV is also 
 found in both observations.
We discuss these results and their physical implications.

\keywords{stars: individual: GX 17+2 --- stars: neutron ---  X-rays: binaries --- accretion, accretion disks}
}

\maketitle

\section{Introduction}
\label{introduction}

 Z sources form the brightest class of galactic Low Mass \mbox{X-ray} Binaries
 (LMXBs), with \mbox{X-ray} luminosities typically in the range
 $\sim$ 0.5--1 times the Eddington luminosity (where $L_{\rm Edd} \sim$  1.5 $M/M_{\odot}\times 10^{38}$ erg s$^{-1}$).
 Their name arises from the pattern they trace in the colour-colour
 or  hardness-intensity diagram (CD and HID, respectively) on typical time scales of days. The three regions which characterize the 
 Z-track, from the top, are the horizontal branch (HB), the normal branch (NB) 
and the flaring
 branch (FB). 
  Multifrequency observations of \mbox{Sco X-1}  (Vrtilek et al. 1991a; Hertz et al. 1992; Augusteijn et al. 1992)
and Cyg X-2 (van Paradaijs et al. 1990; Vrtilek et al 1990; Hasinger et al. 1990) have shown that the accretion rate \deltam\ increasing as the sources move in the direction HB $\rightarrow$ NB $\rightarrow$ FB is consistent with UV observations and some QPO models.
Several two-component models have been proposed to fit the low energy ($\la 20$~keV)
 \mbox{X-ray} spectra of Z sources, such as a simple or Comptonized blackbody (\bb)
 plus a multicolour disk blackbody ({\sc dbb}, the ``Eastern Model'',  Mitsuda et al. 1984, 1989), a \bb\ plus an unsaturated  Comptonization spectrum ($F(E) \propto E^{-\Gamma} \exp{(-E/E_c)}$, the ``Western Model'', White et al. 1986, 1988) and, more recently,
  a \dbb\ plus the Comptonization model worked out by Titarchuk (1994, 
 \comptt\ in \xspec, Piraino et al. 2002; Di Salvo et al. 2002).
 In recent years, transient  hard \mbox{X-ray} tails extending up to 
100~keV have been discovered  from the classical Z sources (GX 5-1, Asai et al. 1994; 
 Cyg X-2, Frontera et al. 1998; Di Salvo et al. 2002; \mbox{GX 17+2}, Di Salvo et 
 al. 2000, hereafter DS2000; Sco X-1, D'Amico et al. 2001; GX 349+2, Di Salvo 
 et al. 2001) and from the peculiar Z source Cir X-1 (Iaria et al. 2001). All these 
hard tails have been fit with power-laws (\pl, $F(E) \propto E^{-\Gamma}$) with photon index 
 $\Gamma$ ranging from $\sim -1$  (\mbox{Sco X-1}, when the source was in the FB) 
to $\sim 3.3$ (Cir X-1). 
The time behaviour of the hard tail is complex and still not  well understood.
 DS2000 observed a strong hard tail in \mbox{GX 17+2} when the
source was on the left part of the HB; as the source moved on the right  across the branch,
an independent fit of the photon index and \pl\ normalization was not possible, and fixing the photon index at 2.7 yielded 
an intensity of the \pl\ that lowered (chance probability of F-test $\sim 10^{-4}$) until
it disappeared at the HB-NB apex.
In the case of Cyg X-2 (Di Salvo et al. 2002) the hard tail was observed when
the source was in the HB, while in a later observation, when the source was
in the NB, no evidence of hard \mbox{X-ray} emission was found.
The hard tail of GX 5-1 (Asai et al. 1994) was detected when the source was in the NB, and
decreased as the source moved to the FB.
If \deltam\ actually increases from HB to FB, these results could be indicative of the presence of a anti-correlation between  hard \mbox{X-ray} emission and accretion rate \deltam.
However, the results of D'Amico et al.  (2001) contrast with this scenario,
as in \mbox{Sco X-1} the hard tail was observed in all branches without any apparent correlation with
the source position on the CD.

\subsection{GX~17+2}
\label{GX 17+2}

\mbox{GX 17+2} is one of strongest Galactic \mbox{X-ray} sources, and was detected
 in the sixties (e.g., Friedman et al. 1967).
Simultaneous \mbox{X-ray} and radio observations, performed by White et al. (1978),
did not show the presence of significant radio emission, but revealed correlated 
\mbox{X-ray} intensity and spectral variations similar to those already observed from 
\mbox{Sco X-1}. 
Observation of the source with the  Gas Scintillation Proportional
 Counter (2--20 keV)  on-board {\it EXOSAT} (White et al. 1986) provided an accurate
determination of the source continuum spectrum, which was fit in the context of the
 Western model with best fit parameters $\ktbb \sim 1$ keV, radius of the {\sc bb}
emitting sphere $\rbb \sim 15$ km, 
$\Gamma \sim 0.6$ and $E_c \sim 4.5$ keV. Also an emission line at $\sim 6.5$ keV 
with  equivalent width (EW) of $\sim$110~eV was detected. 
The data analysis of the same observation, but with 
the  {\rm Medium Energy Proportional Counter Array}, showed that
the 1--30 keV source spectrum  could be fit with a \bb\ 
($\ktbb \sim 1.2$ keV) plus the Comptonization model proposed by Sunyaev \& 
Titarchuk (1980), with electron temperature $\kte \sim 3$ keV and optical
depth of the electron cloud of $\tau \sim 13$ (White et al. 1988).
The column density along the source direction was determined by Vrtilek et al. (1991b) 
with {\it Einstein} (0.2--20 keV), who found a value of $N_{\rm H} 
\sim 1.5 \times 10^{22}$~cm$^{-2}$, assuming a simple \bb\ ($kT_{bb} \sim 1.5$ keV)
as best fit spectrum.
Hoshi \& Asaoka (1993) observed the source with {\it Ginga} in both the HB and 
the NB. By fitting the 1--20 keV \mbox{X-ray} spectrum with the first version of the Eastern model 
({\sc bb + dbb}), these authors found that the inner disk temperature and the projected radius 
of the {\sc dbb} did not change  significantly from the NB to the 
HB, while the \bb\ temperature gradually  increased
from to 2.1 keV to 2.7 keV.
DS2000 analyzed the 0.1--100 keV \mbox{X-ray} spectrum of \mbox{GX 17+2} with \sax, discovering a transient hard \mbox{X-ray} tail in addition to the  persistent continuum.
The latter was well described by a \bb\ plus \comptt\ model, with best fit parameters 
$\ktbb \sim$ 0.6~keV, $\rbb \sim 40$ km, temperature of the seed photons 
to be Comptonized $kT_{s} \sim 1$~keV, electron temperature $kT_e \sim$3~keV and  
optical depth of the Comptonizing cloud $\tau \sim 10$, assuming a spherical geometry.
 Additionally, they 
found evidence for an emission line with energy centroid $E_{\rm l} \sim 6.7$ keV and 
EW $\sim 35$~eV, and an absorption edge with $E_{\rm edge} \sim 8.6$ keV
 and optical depth $\tau_{\rm aedge} \sim 3\times 10^{-2}$.
Type I \mbox{X-ray} bursts from \mbox{GX 17+2} were observed with the {\it Einstein}
 (Kahn \& Grindlay 1984), {\it HAKUCHO} (Tawara et al. 1984)
 and {\it EXOSAT} (Sztajno et al. 1986) satellites.

Despite the \mbox{X-ray} brightness, the optical counterpart of \mbox{GX 17+2} has not yet
been unambiguously identified. The difficulties arise mainly because of the position 
of the source near the Galactic center ($l = 16^{\circ}.4, b=1^{\circ}.3$), 
where the chance of superposition to unrelated field objects is very high.

In this paper we report on the results obtained from two \sax\ observations
of \mbox{GX 17+2}, performed before the October 1999 observation reported by DS2000.
In Sect.~\ref{observations} we describe the observations and data analysis, 
 in Sect.~\ref{results} we present the results of the
 timing and spectral analysis, in Sect.~\ref{discussion} we discuss
 the results and in Sect.~\ref{conclusions} we draw our conclusions.

\section{Observations and data analysis}
\label{observations}

\begin{table*}[!th]
\caption[]{Log of the two observations of GX 17+2 with the on-source
 exposure time for each NFI.}
\begin{center}
\begin{tabular}{cclcccccccc}
\hline
\hline
& Obs.& \multicolumn{1}{c}{Start Time (UT)} & End Time (UT) &
LECS &  MECS& HPGSPC & PDS & 1.8--10 keV MECS  & 13-200 keV PDS\\
 & & & &ks &ks &ks &ks & count rate (s$^{-1}$) & count rate (s$^{-1}$)\\
\hline
 & 1 & 1997 Apr 03 00:27:52 & 1997 Apr 03 06:00:06  & 2.3 & 10.7 & 4.1 & 3.4 & 176 & 35\\
 & 2 & 1997 Apr 21 02:29:00  & 1997 Apr 21 06:13:55  & 1.7 &  6.5 & 2.4 & 2.3 & 190 & 24\\
\hline
\noalign{\vskip 0cm}
\label{tab:log}
\end{tabular}
\end{center}
\end{table*}

\begin{figure*}
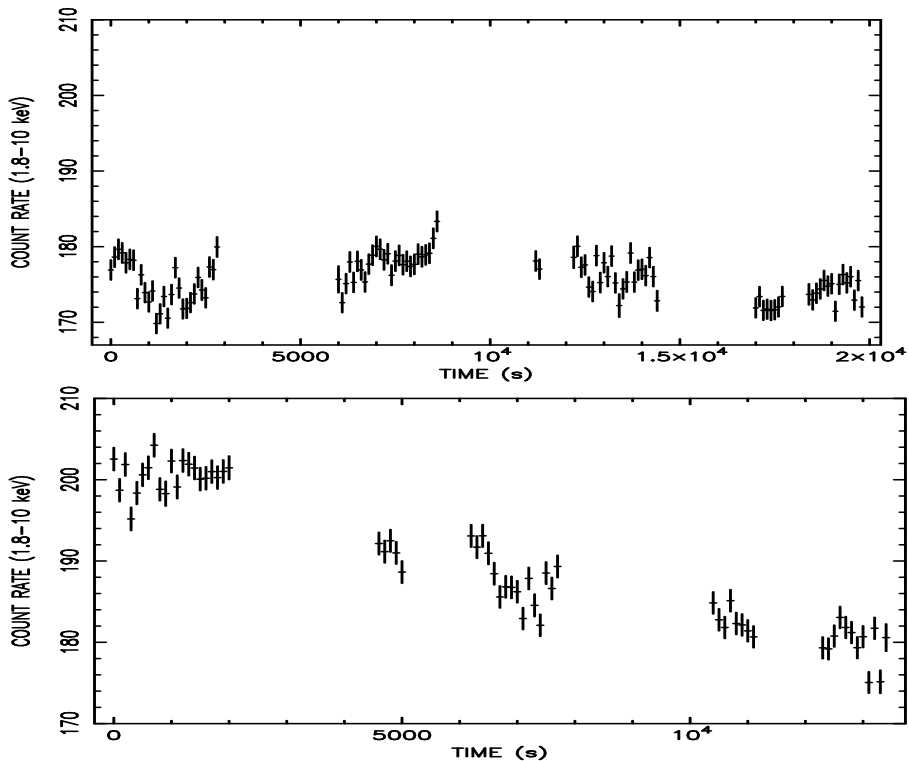

\begin{center}
\epsfig{figure=1567_fig1.ps,height=12cm,width=5cm, angle=-90}
 \vspace{1cm}
\epsfig{figure=1567_fig2.ps,height=12cm,width=5cm, angle=-90}
\end{center}
\vspace{-0.1cm}
\caption[]{MECS 1.8--10 keV light curves of the first ({\it upper panel}) and
 second ({\it lower panel}) observation of \mbox{GX 17+2}. The bin size in both cases
 is 100 s. Time is  given in seconds from the beginning of each observation (see Table \ref{tab:log}).}
\label{f:light_curves}
\end{figure*}

We observed the source twice, on 1997 April 3 and 21, 
with the  Narrow Field Instruments (NFIs) on board \sax\ (Boella et al. 1997a).
The NFIs include a Low-Energy Concentrator Spectrometer (LECS,
0.1--10~keV; Parmar et al. 1997), three Medium-Energy Concentrator
Spectrometers (MECS, 1.5--10~keV; Boella et al. 1997b), a High Pressure
Gas Scintillation Proportional Counter (HPGSPC, 4--120~keV; Manzo et al.
1997), and a Phoswich Detection System (PDS, 15--300~keV; Frontera et
al. 1997). Both observations were performed with all three MECS units  
(MECS unit 1 failed on 1997 May 6). 
Table \ref{tab:log} reports the observations log along with the on-source
exposure times of each NFI and the mean 1.8--10 keV  and 13--200 keV count rate measured by
all the three MECS  units and  by the four PDS units, respectively.
During all pointings the  NFIs worked nominally and the source was
detected by all of them. Good data were selected from intervals when the NFIs 
elevation angle was above the Earth limb by at least $5^{\circ}$ and, for the LECS, 
during the dark earth.
The SAXDAS 2.0.0 data analysis package was used for 
processing  the LECS, MECS and HPGSPC data, while the PDS data reduction was
performed using the XAS package v2.1 (Chiappetti \& Dal Fiume 1997).
The LECS and MECS spectra were extracted from a region of 4\arcmin\ radius
centered on the source position. The background spectrum for the LECS and the MECS
was determined using the standard blank field spectra as reported
by Fiore et al. (1999), while for the HPGSPC and the PDS the background was
estimated with the rocking collimator technique.
The  spectra of all NFIs were rebinned to reduce the oversampling in 
the energy resolution and to have a minimum of 20 counts in each bin, in order to reliably use the 
$\chi^2$ statistics for spectral fitting. 
 For each instrument we performed the spectral and temporal analysis in the 
energy range where the response function is well determined; on the basis of
this extension,the energy range is 0.4--4.0~keV for LECS, 1.8--10~keV for MECS and 8--30~keV for HPGSPC, while for PDS the range is 13--120~keV in the first observation and
13-40 keV in the second one as the source was not detected beyond 40 keV.
We performed spectral analysis over four different regions, named I, II, III and IV,
according to the  1.8--10 keV MECS count rate being $\la$ 180 \s1, between $\sim$ 180 and $\sim$ 185 \s1, between $\sim$ 185 and $\sim$ 195 \s1 and $\ga$ 195 \s1 (see Fig. \ref{f:light_curves}).
We note that region I is entirely covered by the points of the first
observation  while the other regions are covered by the second observation.
Only few points of the second observation fall in region I, so we 
could not extract statistically significant spectral information 
in this region during the second observation. For the latter we then extracted both
the average spectrum and the separated spectra for regions
II, III and IV.

We  used the  \xspec\ v11.2.0 package to fit the multi-instrument energy spectra.
In the broad band fits, normalization factors were applied to LECS, HPGSPC 
and PDS spectra following the cross-calibration tests between them and the MECS,
constraining them within the allowed ranges (Fiore et al. 1999). 
Interstellar photoelectric absorption, modeled using the cross sections 
implemented in \xspec\ and with solar abundances given by  Anders \& Ebihara (1982), 
was included in all  spectral models used.
Uncertainties in the parameters values 
obtained from the spectral fits are single
parameter errors at 90\% confidence level, while upper limits
are reported at 2$\sigma$ level.
For the evaluation of the source luminosity we have assumed a distance
of 7.5 kpc (Penninx et al. 1988)

\section{Results}
\label{results}

\subsection{Temporal properties}
\label{temporal_properties}

\begin{figure*}
\begin{center}
\epsfig{figure=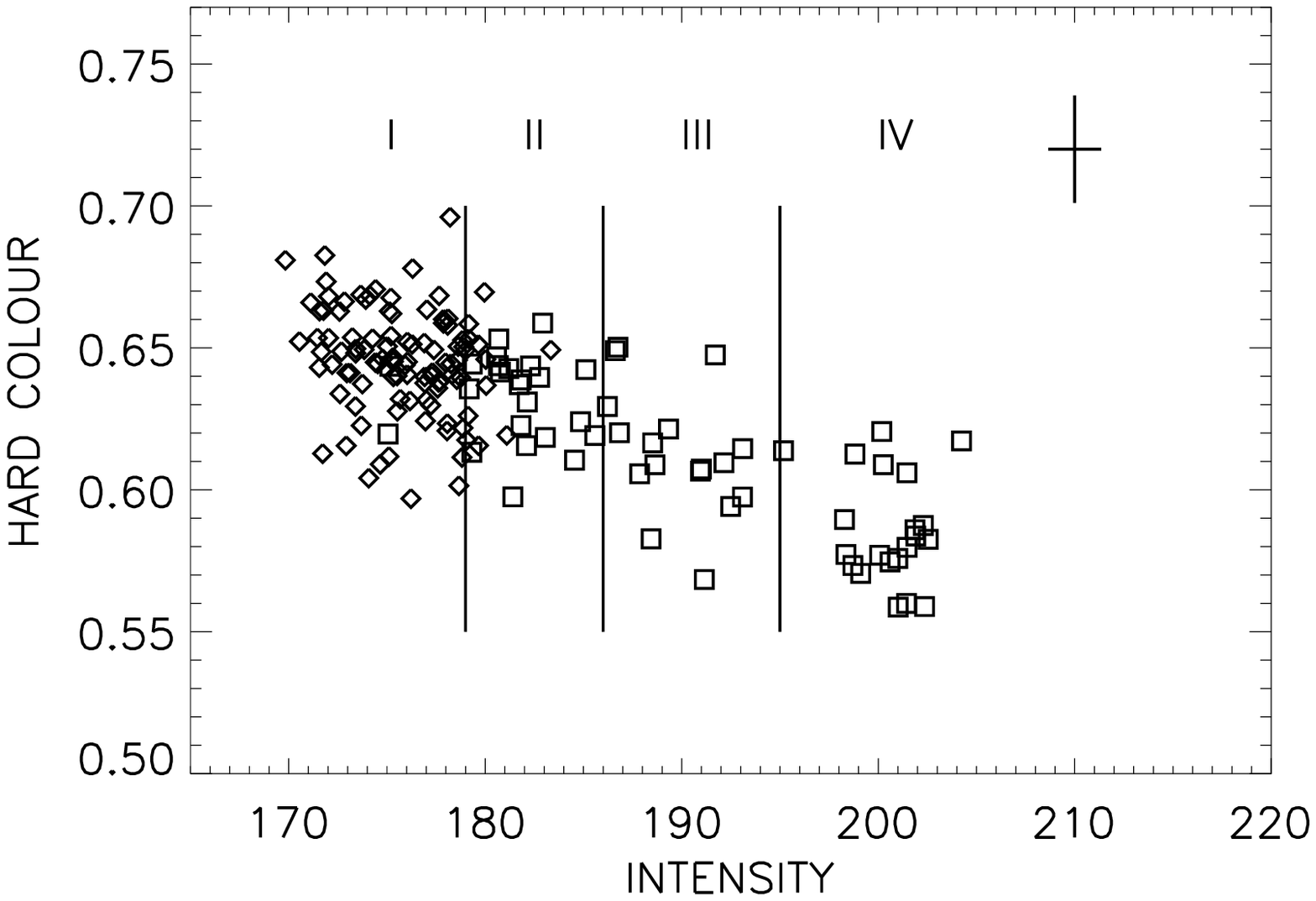, height=7cm, width=7cm}
\epsfig{figure=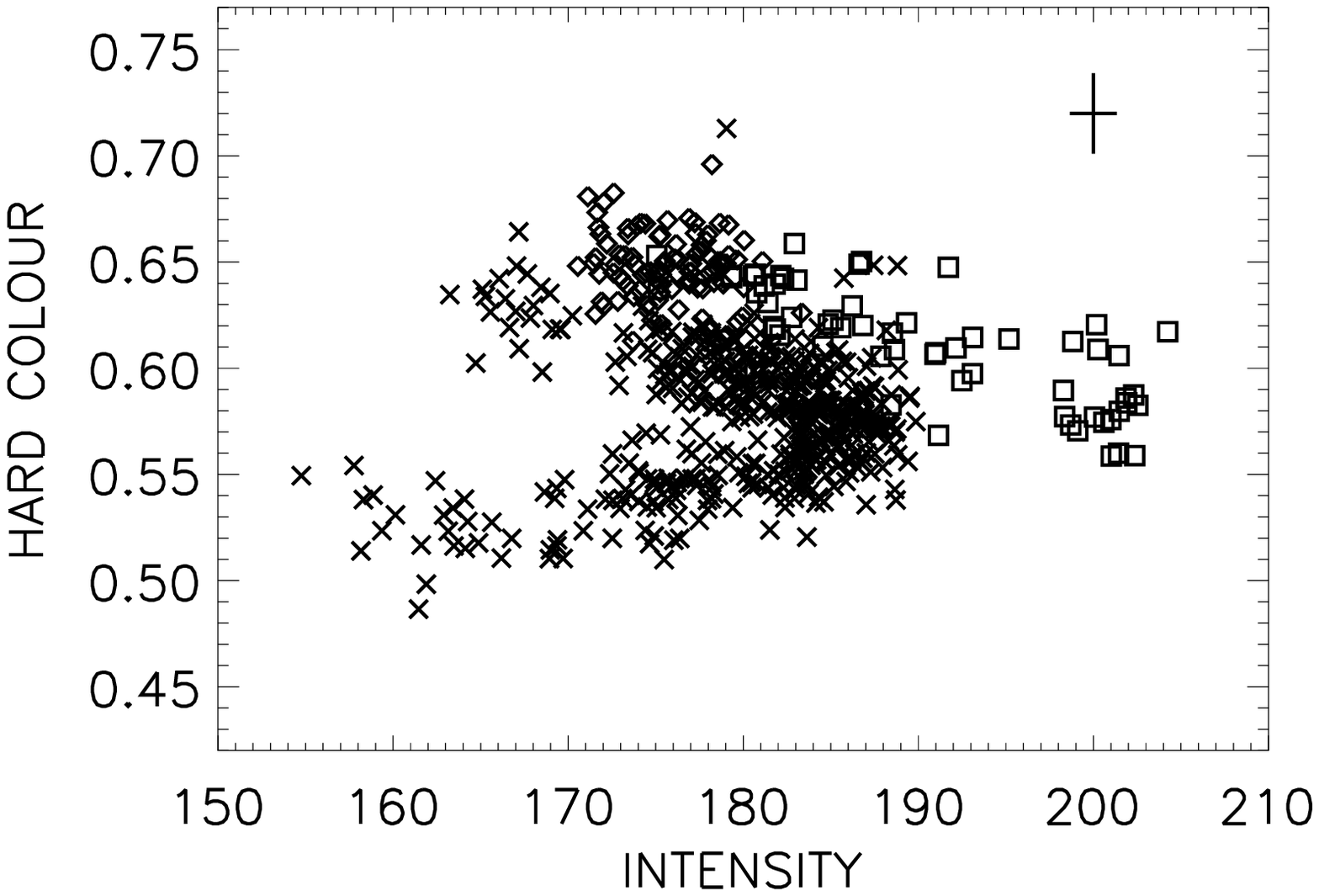, height=7cm, width=7cm}
\end{center}  
\vspace{-0.1cm}
\caption[]{{\it Left panel}: HID of the two \sax\ observations of \mbox{GX 17+2}. 
 {\it Diamonds}: first observation. {\it Open squares}: second observation. 
 The four  count rate regions are also shown. Each point corresponds to a bin size of 100 s. 
 {\it Right panel}: source HID of both
our observations (same symbols as in the left panel) and that performed by DS2000
 ({\it crosses}). The bin size of the 1997 October data points is 400 s.
 A typical error bar is shown on the top right of each panel.}
 \label{f:diagrams}
\end{figure*}

The source 1.8--10 keV light curve of the two observations is shown in 
Fig.~\ref{f:light_curves}. A flux variability on time scales of hours
is apparent and is confirmed by the fit with a constant function, which
gives \chiq/dof = 436/104.
Unfortunately, the presence of several gaps in the light curve prevents us
to investigate whether the source shows periodic flux modulations.
The HID of the two observations,  obtained defining as intensity the 1.8--10 keV MECS count rate  
and as hard colour the count rate ratio in the 6.5--10 keV and 
5--6.5 keV energy bands, respectively, is shown in Fig. ~\ref{f:diagrams}.
As it can be seen, the data points of the first observation are all clustered in 
the left side of the diagram, while those of the second observation, 
reflecting the source intensity decrease (from $\sim$ 200 \s1 to $\sim$ 180 \s1, see Fig.~\ref{f:light_curves}), trace an extended path  (regions II, III and IV) that, 
on the left side, smoothly merges with the data points of the first observation. 
However from the shown HID, it is difficult to 
establish which branch of the Z-track our data correspond to.
To solve this issue, we plot in the lower panel of Fig.~\ref{f:diagrams}  both our 
data points and those obtained with \sax\ by DS2000, when the source clearly traced
the HB and NB.  As DS2000 adopted different energy ranges to produce the HID,
we used the MECS event files of their observations from the \sax\ public data archive to derive 
a new HID in the same energy ranges adopted by us. From Fig.~\ref{f:diagrams} it is possible to see that 
\mbox{GX 17+2}, during our observations, traced almost the entire HB  albeit in a slightly shifted position
with respect to that observed by DS2000. Secular shifts of the Z pattern in the HID of \mbox{GX 17+2} have been
previously observed to occur on time scales of several months by {\it RXTE} (Homan et al. 2002) and {\it EXOSAT} (Kuulkers et al. 1997). On the others hand, both {\it RXTE} and {\it EXOSAT} observations revealed strong
stability (within $\sim 2\%$) of the hard colour, and this confirms the HB nature of the source during
our observations.


We also investigated the source erratic time variability in the 1.8--10 keV energy
band (MECS data),  by deriving the power spectral density (PSD) in the $8 \times 10^{-4}$--50 Hz frequency range, obtained  by averaging  data intervals of 131072 points with a binning time of $10^{-2}$~s,
 for the first and second observation respectively.
 Unfortunately, because of the low collecting area of the telescopes, the statistical quality
of the derived PSD did not allow us  to resolve the broad band noise components typical
of Z sources or to detect QPOs observed with other satellites ({\it EXOSAT}, Hasinger \& van der Klis 1990;{\it RXTE}, Homan et al. 2002).
We thus only derived the integrated root mean square 
fractional variation of the  1.8--10 keV flux  which is 7.1\% $\pm$ 1.3\% in the first observation and  6.3\% $\pm$ 2.4\% in the second one.


\begin{figure*}
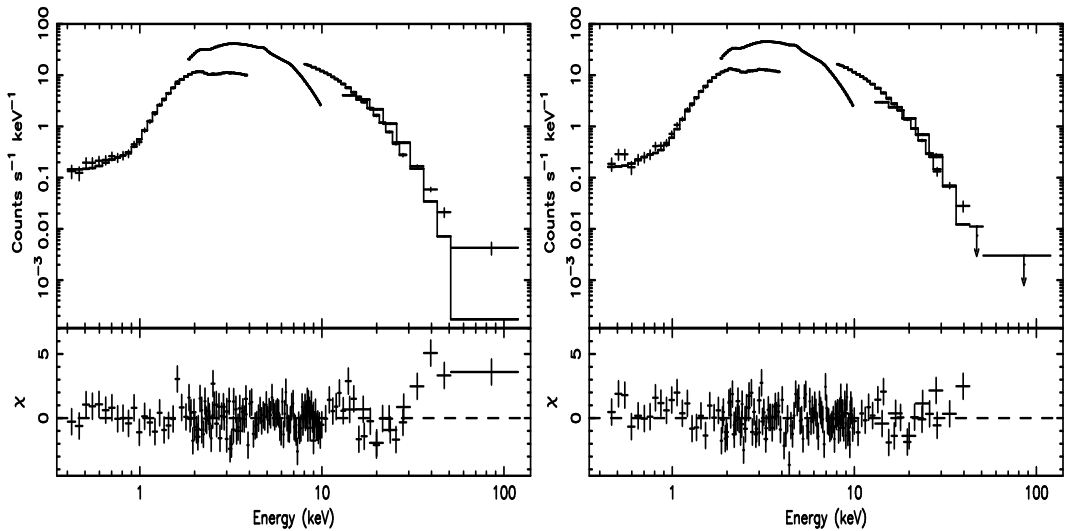

\begin{center}

\epsfig{figure=1567_fig5.ps, height=7cm, width=7cm, angle=-90}
\epsfig{figure=1567_fig6.ps, height=7cm, width=7cm, angle=-90}

\end{center}  
\vspace{0.4cm}
\caption[]{Count rate spectra of the first observation
 ({\it left panel}) and of the second observation
 ({\it right panel}) of \mbox{GX 17+2} and best fit folded model 
  \wabs(\bb\ + \comptt\ + \gaussian) with $\ktbb < \kts$.
In the first observation, an absorption edge was also included.
 Below each panel are shown the residuals to the model in units of $\sigma$.
The last two PDS points of the second observation have S/N $<$ 3,
so they were not included in the fit and are reported as 2$\sigma$ upper
limits.
 Similar results are obtained for the same model but with $\ktbb > \kts$.}
\label{folded_comptt}
\end{figure*}

\begin{figure*}
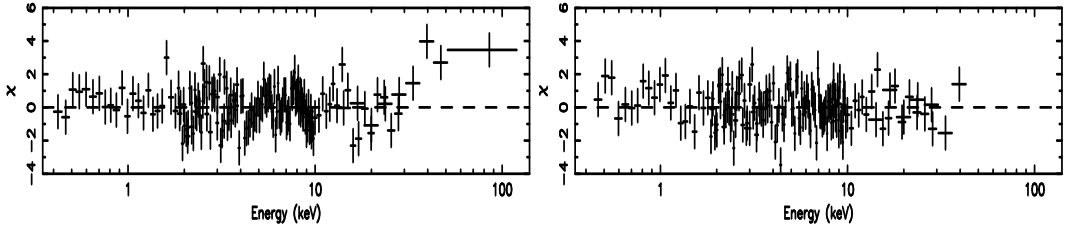

\begin{center}
\epsfig{figure=1567_fig7.ps, height=7cm, width=3cm, angle=-90}
\epsfig{figure=1567_fig8.ps, height=7cm, width=3cm, angle=-90}

\end{center}  

\caption[]{Residuals in units of $\sigma$ to the model \wabs(\bb\ + \eqpair\ + \gaussian), assuming purely thermal Comptonization, for the first ({\it left panel}) and for second ({\it right panel}) observation of \mbox{GX 17+2}.}

\label{folded_eqpair}
\end{figure*}

\begin{table*}
\caption[]{Best fit parameters of the multi-component model \wabs(\bb\ + \comptt\ + \gaussian\ + K-edge + \pl) for the first and second observation of \mbox{GX 17+2}. The two solutions of the model are reported.
 The F-test gives the probability of chance improvement of the \chiq\ for the addition of the \pl. The ratios  $L_{\rm bb}/L_{\rm tot}$  and $L_{\rm pl}/L_{\rm tot}$ are computed in the 0.4--200 keV energy range.} 
\begin{center}
\begin{tabular}{ccccc}
\hline \hline
&  Obs. 1    &  Obs. 2 &   Obs. 1 &  Obs. 2\\
&  I    &  II+III+IV &   I &  II+III+IV\\
\hline
 \multicolumn{1}{c}{Parameter} &\multicolumn{2}{c}{Solution 1}  & \multicolumn{2}{c}{Solution 2} \\
\hline
\noalign{\smallskip}
$N_{\rm H}^{\rm a}$ &$ 2.5^{+0.2}_{-0.3} $ & $2.2^{+0.1}_{-0.2}$ &  $ 2.5^{+0.2}_{-0.3} $   & $ 2.3^{+0.1}_{-0.1} $   \\
$kT_{bb}$ (keV) & $0.66^{+0.04}_{-0.03} $&$0.61^{+0.03}_{-0.03}$ &  $1.79^{+0.10}_{-0.16} $ & $1.38^{+0.05}_{-0.02}$  \\

$R_{bb}$ (km) & 35$^{+4}_{-4}$&  40$^{+4}_{-4}$  &   3.7$^{+0.6}_{-0.6}$   & 7.4$^{+0.5}_{-0.5}$   \\
$kT_{\rm s}$ (keV) & 1.24$^{+0.09}_{-0.08}$ & $1.09^{+0.06}_{-0.04}$  & 0.58$^{+0.02}_{-0.02}$ & $0.55^{+0.01}_{-0.03}$   \\
$kT_{e}$ (keV) & 3.4$^{+0.1}_{-0.1}$ &3.3$^{+0.1}_{-0.1}$& 3.5$^{+0.1}_{-0.1}$ & $3.15^{+0.06}_{-0.04}$   \\
$\tau$ & 11.0$^{+0.4}_{-0.5}$&  10.5$^{+0.4}_{-0.4}$   & 12.0$^{+0.6}_{-0.5}$  & $13.0^{+0.3}_{-0.5}$  \\
$\Gamma $  & 2.8$^{+0.1}_{-0.2}$ & [2.8]   &  2.8$^{+0.1}_{-0.2}$  & [2.8]   \\
$N_{pl}^{\rm b} $  &$ 2.1^{+1.2}_{-1.2}$& 0.7$^{+0.6}_{-0.6}$    &  $ 2.3^{+1.2}_{-1.3}$ &  $1.1^{+0.2}_{-0.6}$ \\

F-test &  $3 \times 10^{-14}$    & 0.11       & $5 \times 10^{-12}$ & 0.03     \\
\hline
\noalign{\smallskip}

$E_{\rm l}$ (keV) &6.7$^{+0.1}_{-0.1}$& 6.7$^{+0.1}_{-0.1}$     & 6.7$^{+0.1}_{-0.1}$  & $6.7^{+0.1}_{-0.1}$    \\
$\sigma_{\rm l}$ (keV) & $0.1^{+0.1}_{-0.1}$& $0.2^{+0.1}_{-0.1}$   &  $0.1^{+0.1}_{-0.1}$ & $0.2^{+0.1}_{-0.1}$    \\
$I_{\rm l}^{c}$ & $3.6^{+1.2}_{-0.8}$& $4.6^{+1.2}_{-1.4}$ &  $3.8^{+1.2}_{-0.8}$& $4.5^{+1.6}_{-1.0}$  \\
$EW_{\rm l}$ (eV)       &  26$^{+9}_{-6}$ & 32$^{+9}_{-9}$  &  27$^{+8}_{-6} $  & $31^{+11}_{-7}$  \\
$ E_{\rm edge}$ (keV) & 8.2$^{+0.4}_{-0.3} $ & [8.2]  &   8.3$^{+0.3}_{-0.2} $   & [8.3]   \\
 $\tau_{\rm edge} (10^{-2}$)& 2.5$^{+1.4}_{-0.7} $ & $<2.0  $  & 3.3$^{+1.1}_{-1.5} $    &$< 1.8 $ \\
\noalign{\smallskip}
\hline

$L_{\rm bb}/L_{tot}$ & 0.16 & 0.17 & 0.10  & 0.15   \\
$L_{\rm pl}/L_{\rm tot}$ & 0.30 & 0.13 &  0.32  & 0.18  \\
$L_{~0.4-30~{\rm keV}}^{\rm d}$  & 1.87 & 1.62 &  1.91  & 1.69   \\

$L_{~30-200~{\rm keV}}^{\rm d}$  & 0.02 &  0.01 &  0.02  & 0.01   \\

${\chi^{2}}$/dof & 143/142 &  188/140 & 140/142   &  189/140  \\

\hline
\noalign{\smallskip}
\multicolumn{5}{l}{$^{\rm a}$ In units of 10$^{22}$ cm$^{-2}$.} \\
\multicolumn{5}{l}{$^{\rm b}$ Photons keV$^{-1}$ cm$^{-2}$ s$^{-1}$ at 1 keV.} \\
\multicolumn{5}{l}{$^{\rm c}$ Total photons in the line in units  of 10$^{-3}$ \cm2 \s1.} \\
\multicolumn{5}{l}{$^{\rm d}$ Unabsorbed  luminosity in units of 10$^{38}$ erg s$^{-1}$.}\\
\noalign{\vskip -0.cm}
\label{t:comptt}
\end{tabular}
\end{center}
\end{table*}


\begin{table*}
\caption[]{Best fit parameters of the multi-component model \wabs(\bb\ + \eqpair\ + \gaussian) for the first and second observation of \mbox{GX 17+2}.
The F-test gives the probability of chance improvement of the \chiq\ switching from thermal to hybrid Comptonization.  The compactness parameters \ls, \lh\  and \lnth\  are  defined in Sect. \ref{eqpair}. The ratio $L_{\rm bb}/L_{\rm tot}$ is computed in the energy range 0.4--200 keV.}
\begin{center}
\begin{tabular}{ccc}

\hline
\hline

Parameter          & Obs. 1		   &	Obs. 2\\
                   & I		   &	II+III+IV\\

\hline

\noalign{\smallskip}

$N_{\rm H}^{\rm a}$ &   2.00$^{+0.05}_{-0.05}$ & 1.97$^{+0.07}_{-0.06}$  \\

$kT_{bb}$ (keV) &   0.58$^{+0.02}_{-0.01}$  &  0.60$^{+0.01}_{-0.02}$     \\

$R_{bb}$ (km) & 33$^{+2}_{-2}$ &  31$^{+3}_{-3}$   \\

$kT_{s}$ (keV) &   0.88$^{+0.14}_{-0.08}$ &   0.95$^{+0.03}_{-0.05}$  \\

$kT_{\rm e}$ (keV) &  3.5  & 3.5    \\

$\ell_{s}$  & 7600$^{+7000}_{-3400}$ &  [7600] \\ 

$\ell_{h}/\ell_{s}$   & 0.87$^{+0.02}_{-0.02}$ & 0.63$^{+0.01}_{-0.03}$    \\ 

$\ell_{nth}/\ell_{h}$ & 0.11$^{+0.02}_{-0.02}$ &  $< 0.04$ \\

$\tau$ &  8.4$^{+1.5}_{-0.7}$  & 8.4$^{+0.1}_{-0.2}$     \\

$p $  & 1.7$^{+1.1}_{-0.6}$ &  [1.7]\\

$\gmin $  &  1.15$^{+1.54}_{-0.15}$   & [1.15] \\

$\gmax $  & [12] & [12] \\

F-test & $ 4 \times 10^{-11}$  &  0.07    \\

\hline
\noalign{\smallskip}

$E_{\rm l}$ (keV) & 6.7$^{+0.1}_{-0.1}$ &6.7$^{+0.1}_{-0.1}$   \\

$\sigma_{\rm l}$ (keV) &   $0.2^{+0.1}_{-0.1}$    & 0.2$^{+0.1}_{-0.1}$   \\

$I_{\rm l}^{b}$ &  $4.4^{+1.3}_{-0.8}$  & $4.8^{+2.2}_{-0.8}$    \\

$EW_{\rm l}$ (eV)  &   32$^{+9}_{-6} $ & 33$^{+15}_{-6} $     \\

\noalign{\smallskip}
\hline

$L_{\rm bb}/L_{\rm tot}$ & 0.11 &  0.11  \\

$L_{~0.4-30~{\rm keV}}^{\rm c}$ & 1.44 & 1.46 \\

$L_{~30-200~{\rm keV}}^{\rm c}$ & 0.019 &  0.006 \\

${\chi^{2}}$/dof & 143/142 &  185/141   \\

\hline
\noalign{\smallskip}

\multicolumn{3}{l}{$^{\rm a}$ In units of 10$^{22}$ cm$^{-2}$.} \\
\multicolumn{3}{l}{$^{\rm b}$ Total photons in the line in units of 10$^{-3}$ \cm2 \s1.} \\
\multicolumn{3}{l}{$^{\rm c}$ Unabsorbed luminosity in units of 10$^{38}$ erg s$^{-1}$.}\\

\noalign{\vskip -0.cm}
\label{t:hybrid}
\end{tabular}
\end{center}
\end{table*}


\begin{table*}
\caption[]{Same as in Table \ref{t:hybrid} but for regions II, III and IV of the second observation as reported in Fig. \ref{f:diagrams} and assuming thermal Comptonization.}
\begin{center}
\begin{tabular}{cccc}
\hline
\hline

Parameter          & II   &	III  &  IV\\

\hline

\noalign{\smallskip}

$N_{\rm H}^{\rm a}$ &   2.03$^{+0.11}_{-0.05}$ & 2.05$^{+0.10}_{-0.13}$   & 1.93$^{+0.18}_{-0.14}$\\

$kT_{bb}$ (keV) &   0.61$^{+0.01}_{-0.03}$  &  0.53$^{+0.09}_{-0.06}$ &  0.55$^{+0.08}_{-0.06}$    \\

$R_{bb}$ (km) & 37$^{+3}_{-3}$ &  32$^{+10}_{-10}$    &  29$^{+9}_{-9}$   \\

$kT_{s}$ (keV) &   1.17$^{+0.02}_{-0.07}$ &   0.85$^{+0.05}_{-0.08}$ &   0.90$^{+0.10}_{-0.04}$ \\

$kT_{\rm e}$ (keV) &  4.0  & 3.5  & 3.1  \\

$\ell_{s}$  & [7600] &  [7600]  & [7600]\\ 

$\ell_{h}/\ell_{s}$   & 0.47$^{+0.01}_{-0.03}$ & 0.71$^{+0.04}_{-0.03}$  & 0.59$^{+0.02}_{-0.02}$   \\

$\tau$ &  6.7$^{+0.2}_{-0.5}$  & 8.6$^{+0.4}_{-0.3}$  & 8.9$^{+0.3}_{-0.2}$     \\

\hline
\noalign{\smallskip}

$E_{l}$ (keV) & 6.6$^{+0.1}_{-0.1}$ &6.7$^{+0.2}_{-0.2}$  &6.8$^{+0.1}_{-0.1}$ \\

$\sigma_{l}$ (keV) &   $0.2^{+0.4}_{-0.2}$    & 0.3$^{+0.3}_{-0.2}$ & 0.1$^{+0.2}_{-0.1}$  \\

$I_{\rm l}^{b}$ &  $5.0^{+3.6}_{-2.3}$  & $6.5^{+4.8}_{-3.1}$   & $5.3^{+2.4}_{-1.8}$  \\

$EW_{\rm l}$ (eV)  &   35$^{+25}_{-16} $ & 44$^{+33}_{-21} $   & 36$^{+16}_{-12} $    \\

\noalign{\smallskip}
\hline

$L_{\rm bb}/L_{\rm tot}$ & 0.17 &  0.07 & 0.07 \\

$L_{~0.4-30~{\rm keV}}^{\rm c}$ & 1.47 & 1.50 & 1.48\\

$L_{~30-200~{\rm keV}}^{\rm c}$ & 0.01 &  0.007  & 0.004\\

${\chi^{2}}$/dof & 145/142 &  133/142  & 145/142 \\

\hline
\noalign{\smallskip}

\multicolumn{3}{l}{$^{\rm a}$ In units of 10$^{22}$ cm$^{-2}$.} \\
\multicolumn{3}{l}{$^{\rm b}$ Total photons in the line in units of 10$^{-3}$ \cm2 \s1.} \\
\multicolumn{3}{l}{$^{\rm c}$ Unabsorbed luminosity in units of 10$^{38}$ erg s$^{-1}$.}\\

\noalign{\vskip -0.cm}
\label{t:regions}
\end{tabular}
\end{center}
\end{table*}


\begin{figure*}[ht]
\begin{center}
\epsfig{figure=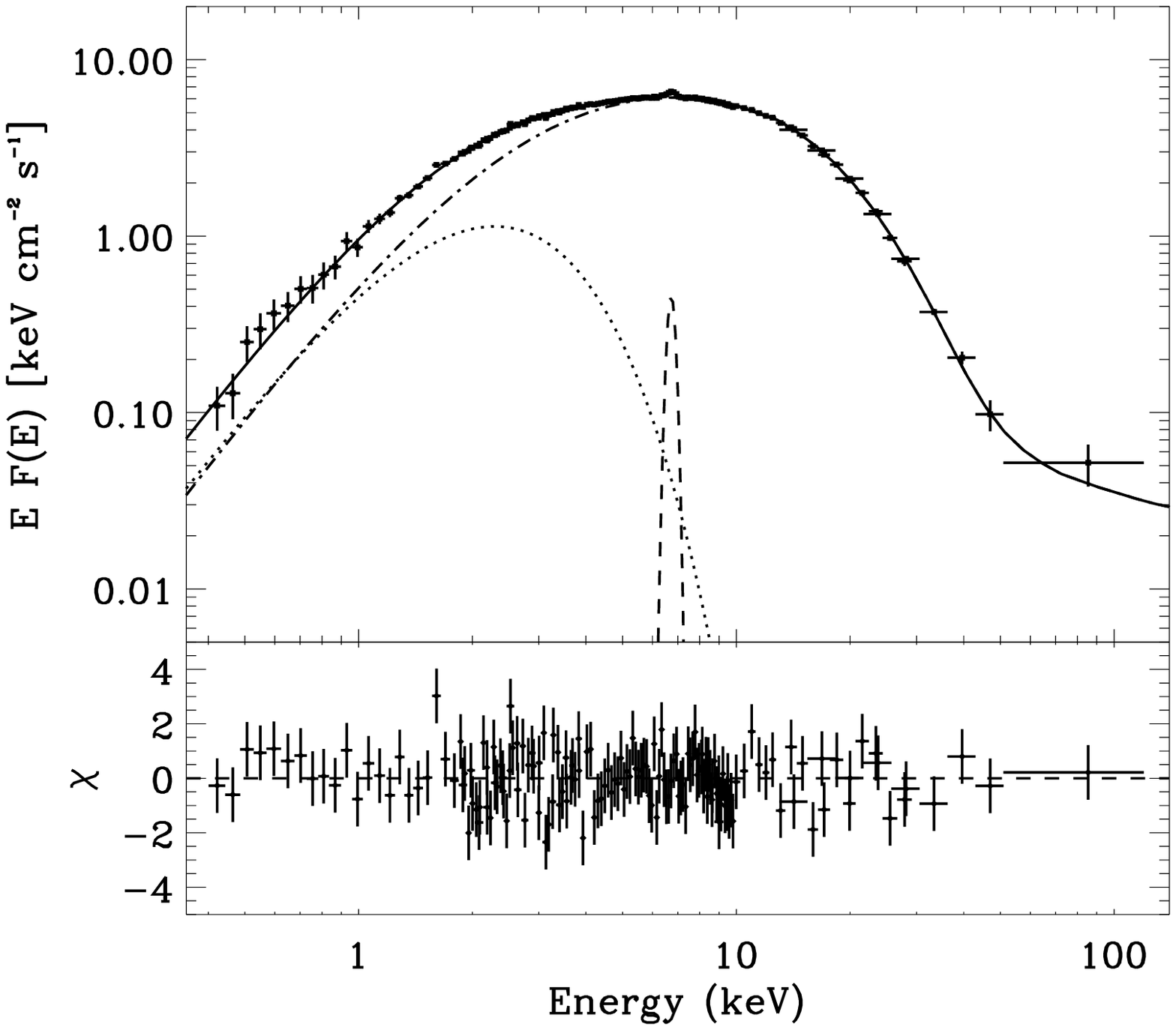, width=7.5cm, height=6cm}
\epsfig{figure=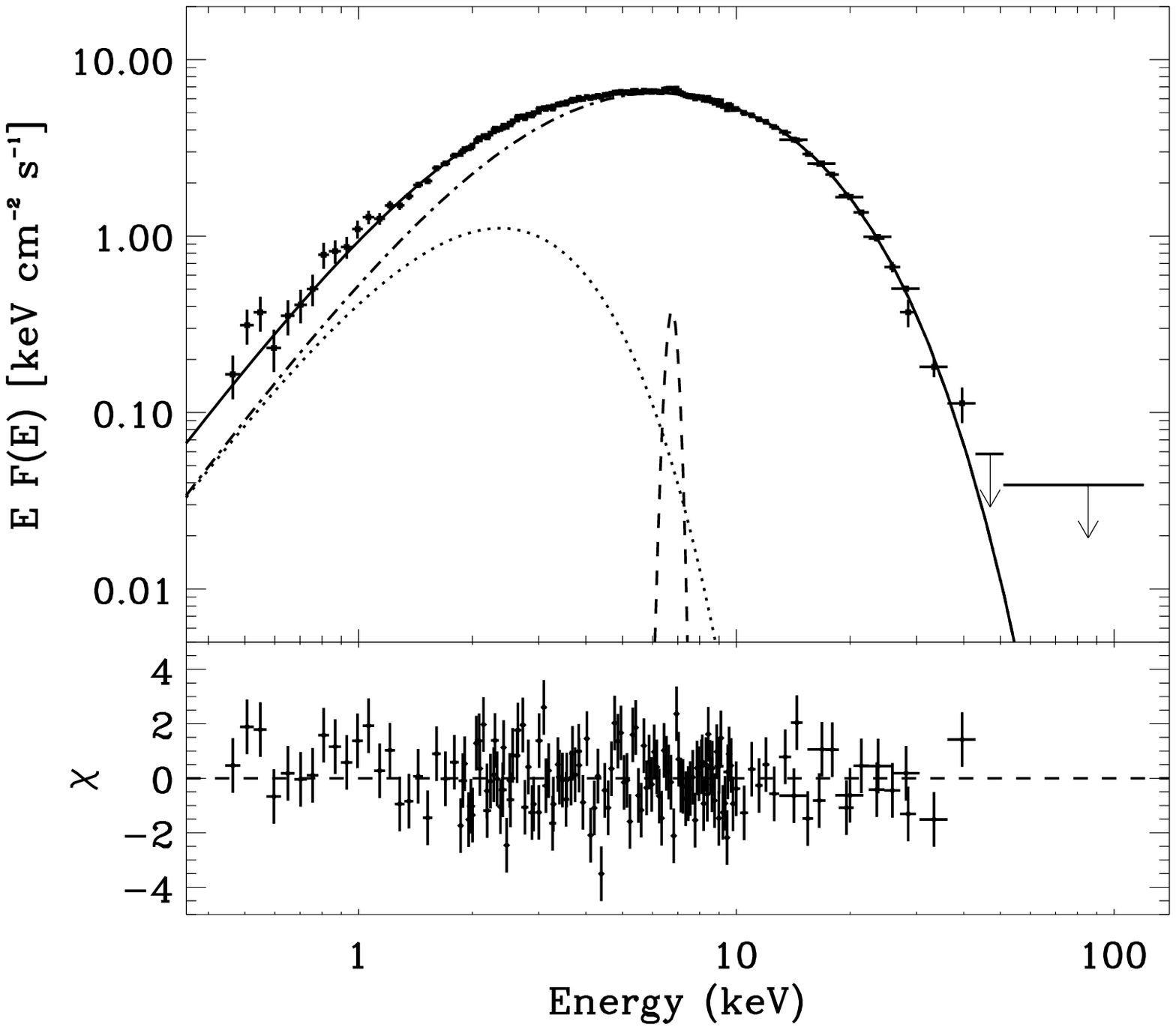, width=7.5cm, height=6cm}
\end{center}
\vspace{0.5cm}
\caption[]{Unabsorbed $EF(E)$ spectrum of the first ({\it left panel}) and of the second ({\it right panel}) 
observation of \mbox{GX 17+2} along with the best fit model \wabs(\bb\ + \eqpair\ + \gaussian). 
A hybrid spectrum of the Comptonizing electrons is assumed. The 2$\sigma$ upper  limit
of the last two PDS points of the second observation is also reported.
Below each  panel, the residuals between the data and the model
in units of $\sigma$ are shown.
Different line styles represent the single components. {\it Dotted}: \bb.
{\it Dotted-dashed}: \eqpair. {\it Dashed}: Gaussian emission line.}
\label{eeuf_eqpair}
\end{figure*}


\subsection{Spectral properties}
\label{s:spectra}

Following the results obtained on \mbox{GX 17+2} by DS2000, we first attempted 
to fit  the source spectra with the {\sc bb + comptt} continuum model plus a fluorescence 
Fe line and an absorption Fe K-edge. The best fit results are shown in
Fig.~\ref{folded_comptt}, where the time-averaged count rate spectra of the 
observations 1 and 2 are shown along with the residuals to 
this model.
As can be seen, for the second observation (Fig.~\ref{folded_comptt}) the model 
describes quite well the data, with no apparent systematic residuals, 
even though the \chiq\ is not satisfactory 
($\chi^2/{\rm dof} = 192/141$). This   could be due to some spectral 
evolution of the source with time (see Sect.~\ref{obs2}). In addition, no significant \mbox{K-edge} is required by the data (see Table \ref{t:comptt}). 
 Unlike the second observation, the above model does not provide a good 
description of the data of the first observation, with a $\chi^2/{\rm dof} = 
222/144$, mainly due to the residuals to the best fit model above 30~keV (see Fig.~\ref{folded_comptt}). It is sufficient to add, as
also done by DS2000, a \pl\ component to the above spectral model to get a 
very good fit ($\chi^2/{\rm dof} = 143/142$). The best fit results are 
shown in Table \ref{t:comptt}.  A feature of the data obtained from both observations
when they are fit with the above model (with or without the \pl) is that two 
equally acceptable solutions for the values of the \bb\ temperature $kT_{bb}$ 
and temperature of the seed photons $kT_s$ are found (see Table \ref{t:comptt}): one with $kT_{bb} < kT_s$ and the other with $kT_{bb} > kT_s$, while the other parameter values remain substantially unchanged within  errors.

Finally, to test the physical hypothesis of the three-layered atmospheric
structure in accretion disks suggested by Zhang et al. (2000)
for the  black hole candidates (BHC) \mbox{GRO J1655--40} and \mbox{GRS 1915+105},
in the first observation we replaced the \pl\ component with a 
further \comptt\ model with higher electron temperature, which we label
as $\kteh$.
Approximating the seed photons temperature of this hotter electrons to the
temperature of the electrons of the underlying warm layer
(an assumption justified by the fact that the Compton parameter
 $y$ of the warm electrons is $\sim$ 3, so that their emerging spectrum
 is not far from a Wien-like), namely keeping $\ktsh=\kte$, does not give any stable solution.
A good fit is obtained by keeping free all the model parameters 
(\chiq/dof = 149/140), but in this case, unlike the results
by Zhang et al. (2000), the seed photons temperature of the hotter
\comptt\ component ($\ktsh \sim 1.3$ keV, with a corresponding $\kteh \sim 75$ keV and $\tau^h \sim 0.05$) is quite different from the electron temperature ($\sim 8$ keV) 
of the warm electrons.

\subsection{A physical model for the hard X-ray tail}
\label{eqpair}

In order to investigate the physical origin of the hard \mbox{X-ray} tail 
detected during the first observation of \mbox{GX 17+2}, we used the Comptonization model 
worked out by Coppi (1999, \eqpair\ in {\sc xspec}), which has the advantage of being
general and can be used to fit the source spectra of both observations.
In this model, non-thermal electrons with energy spectrum $\propto \gamma^{-p}$
(where $\gamma$ is the Lorentz factor) are injected in 
the plasma cloud at a rate $\propto \gamma^{-p}$.
These electrons lose energy because of Compton, Coulomb and
bremsstrahlung processes and thus establish a steady-state distribution.
At high energies (for $\gamma > \gmin$ up to a certain value
of $\gmax$) the distribution is non-thermal (power-law like), but
at low energies  it joins the  Maxwellian distribution and
the final result of this process is a hybrid thermal plus non-thermal 
plasma.
The temperature of the thermal distribution, $\kte$, is self-consistently computed by means
of energy balance and is not a free parameter of the model.
The plasma properties are characterized by the powers $L$ supplied
to the different components.
A compactness parameter is introduced and defined as:

\begin{equation}
\ell \equiv {\sigma_{\rm T} \over m_{\rm e} c^2} {L\over R},
\label{compactness}
\end{equation}

\noindent
where $R$ is the characteristic size of the region and ${\sigma_T}$ the Thomson
cross-section for scattering.
The compactness of the input soft photons is parametrized
by \ls. The power supplied to the Comptonizing electrons is parametrized as
\lh=\lth + \lnth, where \lth\ is the power supplied to heat
thermal particles, while \lnth\ is the power going into accelerating
non-thermal particles. If \lnth=0, then one has a purely
thermal (Maxwellian) distribution. \eqpair\ can also test the presence of Compton 
reflection of the primary photon spectrum  from a neutral or a ionized medium
(Magdziarz \& Zdziarski 1995) allowing for different metal abundances
with respect to the solar ones.

\subsubsection{First observation}

We first assumed that all the power goes in heating the thermal particles
(\lnth/\lh = 0, purely thermal Comptonization) with 
a simple \bb\ spectrum of the seed photons.
We added to \eqpair\ also a direct \bb\ component plus a Fe K fluorescence line in 
the source model, as discussed in  Sect.~\ref{s:spectra}.
This model well describes the data below 30 keV but, as in the case
of \comptt, a systematic excess is again observed in the residuals
above this threshold (see Fig. \ref{folded_eqpair}).
We thus left free to vary the \lnth\ parameter: since the statistics of the data do not allow us to simultaneously constrain $\gmin$,  $\gmax$ and $p$ in the fits, $\gmax$
was frozen at value of 12 (the \chiq\ is poorly sensitive to $\gmax$ in the range from 2 to 50). The chosen value of $\gmax$
takes into account the  energy range in which the excess over thermal Comptonization is located and the temperature of the seed photons found from the fit.
As can be seen from Table \ref{t:hybrid}, the agreement between
the model and the data is excellent.
The probability given by the F-test that the $\chi^2$ reduction obtained from the inclusion of a non-thermal component in the Comptonizing cloud is due to chance is very low 
(see Table \ref{t:hybrid}).
The presence of a Compton reflection component, assuming solar abundances and a low inclination angle $i=30^{\circ}$ (\mbox{GX 17+2} is a \mbox{Sco-like} source, Kuulkers \& van der Klis 1996), 
 is only marginally detected ($\Delta$\chiq $\sim$ 7, probability of chance improvement of $\sim$ 3\%) with best fit value of the covering factor  $\covfact \sim 0.04$.
The ionization parameter value we find is $\xi \sim 800$  erg cm s$^{-1}$.
An important issue to point out is that, unlike \comptt, with \eqpair\ there is not
a double solution as concerns the direct \bb\ and seed photons
temperature values.

Replacing the direct \bb\ component with a \dbb, we still obtain 
an acceptable, albeit a bit worse, result (\chiq/dof = 156/142).
If, on the other hand, we assume a \dbb\ seed photon spectrum
in \eqpair\ plus a direct \bb\ model, the fit is even worse (\chiq/dof = 163/142).

\subsubsection{Second observation}
\label{obs2}

As reported in Sect. \ref{s:spectra}, during the second observation
there was slight evidence of a spectral evolution of the source,
suggested by the somewhat worse fit of the \comptt\ model to the  data, 
in  spite of the fact that no systematic 
residuals to the best fit model are apparent (see Fig.~\ref{folded_comptt}).
This result is confirmed also by using the \eqpair\ model, assuming
purely thermal Comptonization, to fit the whole observation  (\chiq/dof = 186/142).
In this case we also find that the soft compactness \ls\ is unconstrained so, during the fit, 
we fixed it at the best fit value of the first observation. 
In order to put an upper limit on the power supplied in accelerating
non-thermal electrons, we allowed the \lnth/\lh\ parameter to
vary, by keeping $\gmin$ and $p$ frozen at the best fit values
of the first observation. The results are reported in Table \ref{t:hybrid}.
As it can be seen, the power going in accelerating  non-thermal electrons is
consistent with zero and its 2$\sigma$ upper limit is lower than the value 
of the first observation. It is worth pointing out that this
upper limit value on  \lnth/\lh\ is obtained even including in the fit the last two PDS points 
(see  Fig. \ref{folded_comptt}),in order to have the same energy coverage of the first observation.
The  2$\sigma$ upper limit on the  40-120 keV source flux, assuming as best fit model 
the one reported in Table \ref{t:hybrid}, is  $\sim 7 \times 10^{-11}$ erg cm$^{-2}$ s$^{-1}$
(see Fig. \ref{eeuf_eqpair}) a value below that found in the first observation 
($\sim 1.0 \times 10^{-10}$ erg cm$^{-2}$ s$^{-1}$).
We subsequently  performed spectral analysis separately for the three regions of the HID
labeled in Fig. \ref{f:diagrams}; the results are reported in Table \ref{t:regions}.

\section{Discussion}
\label{discussion}

Some relevant results have been derived from our observations of \mbox{GX 17+2} 
with {\it BeppoSAX}.
The persistent continuum \mbox{X-ray} emission of
\mbox{GX 17+2} can be described by several 
 two-component models. When we adopt the \comptt\ model 
plus a \bb, we find that two solutions are possible, which are 
reminiscent of the well known Eastern  (Mitsuda et al. 1984, 1989) 
and Western (White et al. 1986, 1988)  models.
Indeed, looking at the results of  Table~\ref{t:comptt}, for solution
1, we can interpret the $\sim$ 0.6 keV \bb\  as the direct emission from the accretion disk, with Comptonization mainly occurring 
in the hotter boundary  layer ($\kts \sim 1.2$ keV, Eastern Model scenario).
But the same spectrum can be described (solution 2) by a $\sim$ 1.5 keV \bb\
plus a Comptonization region of cooler photons ($\kts \sim 0.6$ keV). In this case
the direct \bb\ emission can be interpreted as coming from the boundary layer,
while Comptonization takes place in the disk (Western model scenario).
From a statistical point of view, we do not find strong reasons for favouring 
one solution over the other; however, the former solution is the only one 
we find when  \comptt\ is replaced by \eqpair.

The soft compactness estimated by \eqpair\ (see Table \ref{t:hybrid}) is
much higher than that reported in the case of BHCs (e.g., Frontera et al. 2001;
\gierlinski\ \& Done 2003); this is however not surprising given that in neutron star systems,
unlike BHCs, the source environment in the inner disk region 
has an additional source of soft luminosity coming from the neutron star.
Apart from this crucial difference, related to the nature of the compact
object, there are some physical characteristics which link in some
way \mbox{GX 17+2} to BHCs.
First, the spectral index of the non-thermal  injected electrons ($p \sim$ 1.7) 
is consistent, within errors, with that found, e.g., in Cyg X-1 (Frontera et al. 2001)
 and \mbox{GRS 1915+105} (Zdziarski et al. 2001).
Moreover, the spectral shape of the source in the $EF(E)$ diagram 
(see Fig. \ref{eeuf_eqpair}) strongly resembles
that of BHCs in the so-called ultrasoft state, when
$L_{\rm X} \sim L_{\rm Edd}$ (e.g., \mbox{GX 339--4}, \mbox{GRS 1915+105}, \mbox{XTE J1550-564}, 
see the review by Zdziarski \& \gierlinski\ 2004), where the peak of the energy
emission occurs below 10 keV.
 The similarity with \mbox{GRS 1915+105}, in particular, results even more
intriguing looking at the \mbox{X-ray} spectral parameters found
by Zdziarski et al. (2001) by fitting the source {\it RXTE/OSSE} 3--600 keV
\mbox{X-ray} spectrum  with \eqpair: the authors indeed found 
$\kte \sim$ 3.6 keV, $\tau \sim$ 4.4, $\kts \sim$ 1.3 keV and
\lnth/\lth $\sim$ 0.1. Only the ratio \lh/\ls was  lower
($\sim$ 0.3) than that found in \mbox{GX 17+2} by us.
As Z sources stably accrete at near-Eddington rate, these similarities 
lead to suggest that Comptonization/acceleration
processes in \mbox{X-ray} binary systems in this phase do not strongly depend on
the presence or not of a solid surface at the inner boundary
of the accretion disk.

An important result of our study concerns the transient high energy  \mbox{X-ray} 
tail. We confirm its presence when the source is in the upper  left HB 
of the Z diagram, as already observed by DS2000.
But, unlike DS2000, who found that the hard \mbox{X-ray} excess 
gradually decreased as the source moved towards right in the HB, 
completely disappearing when the source achieved the NB, 
we find that this disappearance occurs when the source is still 
in the HB, in particular during its transition 
from region I to region II of the HID (Fig.~\ref{f:diagrams}).
The hard \mbox{X-ray} excess is described by a \pl\ with
the same photon index found by DS2000 
($\Gamma \sim$ 2.7, see Table \ref{t:comptt}).
In order to gain some insight on the physical origin of the 
hard \mbox{X-ray} component, we replaced the \comptt\  model with the 
Comptonization model \eqpair\ of Coppi (1999).
In this model, in addition to a thermal population of electrons, 
a non-thermal electron tail can be included.
Assuming a purely Maxwellian distribution of the electron energies, the hard
\mbox{X-ray} excess is still apparent  above the \bb\  plus \eqpair\ model
(see Fig. \ref{folded_eqpair}).
Only assuming a thermal plus non-thermal plasma the \eqpair\ model gives
a good description of the high energy excess.
To our knowledge, this is the first time that the transient hard \mbox{X-ray} tail of
a Z source (see Sect. \ref{introduction}) is described in
terms of hybrid Comptonization.

An important question is whether the two electron  populations (thermal and 
non-thermal) are physically connected. Unfortunately the present data do not 
allow to answer this question. We consider  two cases separately.
If the two electron populations are not physically related, a likely origin 
of the non-thermal  electrons is a relativistic jet escaping the system. 
In recent years, relativistic jets  have revealed  to be a quite common property of LMXBs, providing a link between these systems and AGNs (for a review see, e.g., Fender 2002).
In this thermal plus non-thermal scenario, the Comptonized spectrum 
is produced by a thermal corona  (likely located above the inner disk close 
to the boundary layer) plus a relativistic non-thermal electron jet, 
with the soft seed photons mainly coming from the inner disk. 
The energy of the seed photons ($\sim$ 1 keV, see Table \ref{t:hybrid}) 
is indeed consistent with the value expected for the inner disk temperature
of LMXBs hosting a neutron star (see, e.g., the review by Barret 2001).
Thus we could tentatively identify the decrease of the power
supplied to non-thermal electrons (\lnth/\lh\ drops from $\sim$ 0.1
to less than  $\sim$ 0.04 from the first to the second observation),
as a switching-off of the mechanism producing the jet.

On the other hand, if the two electron populations are physically related, a possible
origin of the non-thermal component is in processes of magnetohydrodynamic  
turbulence occurring in the corona (e.g., magnetic reconnection), in analogy 
with what  observed in the solar corona (Crosby et al. 1998, 
see also the review by Coppi 1999). In this case particle acceleration cannot be 100\% efficient 
in the production of non-thermal high energy particles 
(e.g., Zdziarski et al. 1993). 
This can occur  only above a certain momentum 
threshold  when the acceleration time scale for a particle becomes significantly 
shorter than its corresponding thermalization (or energy-exchange) time scale.
Below this threshold particles are tightly coupled and their acceleration
goes mainly in a bulk heating of the thermal plasma component.

We have also explored the physical scenario proposed by Zhang et al. (2000) for \mbox{GRO J1655--40} and
\mbox{GRS 1915+105}, in which a much hotter, optically thin thermal corona is located
above a warm layer with lower temperature and high opacity.
This scenario seems not suitable for \mbox{GX 17+2}. In fact, in this hypothesis, one would
expect that the seed photons of the hotter layer are those coming from
the underlying one, and thus should have a temperature similar
to that of the colder electrons ($\ktew \sim$ 3--5 keV).
This is indeed what Zhang et al. (2000) found with {\it ASCA/RXTE} data
in the case of   \mbox{GRO J1655--40}  and \mbox{GRS 1915+105}.
However, in our case $\ktsh$ significantly differs from $\ktew$ ($\sim 1.3$ keV
and $\sim$ 8 keV, respectively), thus ruling out the
three-layered accretion hypothesis.

In addition to the origin of the hard \mbox{X-ray} emission, its correlation
with the position of Z sources in the CD/HID is still
unclear. 
As discussed in Sect. \ref{introduction}, the results obtained
from \mbox{GX 17+2}, \mbox{Cyg X-2} and \mbox{GX 5-1} (but not, however,  from \mbox{Sco X-1}), could suggest the possibility of an anti-correlation between presence of hard \mbox{X-ray} emission and accretion rate \deltam\ in Z sources, in a similar fashion to what observed both in atoll sources and BHCs, where the hard spectra are 
usually observed during the low  inferred \deltam\ states (e.g., Barret \& Olive 2002; McClintock \& Remillard 2004).
Our results on \mbox{GX 17+2} do not contrast with this possibility but,
when compared with those found by DS2000, they point  to the presence 
of a threshold, that we thus tentatively identify with a threshold in \deltam, above which the hard tail disappears.
A definite answer to this suggestion could come from very sensitive monitoring of the source at high energies in region I of Fig. \ref{f:diagrams} during the second observation, but this was not possible because of the too few points falling in that region.

 An important issue is what we mean by \deltam: it appears evident that this \deltam\ is not the one that drives the \mbox{X-ray} luminosity, since the latter does not change along the track (see Tables \ref{t:hybrid}  and \ref{t:regions}).
Thus, if some \deltam\ increases along the Z-track, it is not correlated with the \mbox{X-ray} luminosity: an intriguing way out to this problem was proposed by van der Klis (2001) to explain the ``parallel tracks'' phenomenon. He suggested that what does change along the track is not the prompt disk accretion rate $\dot{M}_{\rm d}(t)$ but rather the ratio  $\eta=\dot{M}_{\rm d}(t)/\langle \dot{M}_{\rm d}(t)\rangle$, where $\langle \dot{M}_{\rm d}(t)\rangle$ is the long-term $\dot{M}_{\rm d}(t)$ average.
The luminosity is supposed to be related to the sum of $\dot{M}_{\rm d}(t)$ plus  $\langle \dot{M}_{\rm d}(t)\rangle$ ($L_{\rm X} \propto \dot{M}_{\rm d}(t) + \alpha \langle \dot{M}_{\rm d}(t) \rangle$) while the curve length S across the track is a measure of $\eta$.
The threshold above which we see the disappearance of the hard tail could thus be a threshold in $\dot{M}_{\rm d}(t)/\langle \dot{M}_{\rm d}(t) \rangle$.
It is quite difficult to establish whether this hypothesis is correct.
In the positive, one has to explain what produces the hard tail: the physical process should be strongly dependent on what happens in the disk.
When the hard \mbox{X-ray} component  is no longer detected (regions II, III and IV of the HID), we observe a slight decrease of the temperature of the scattering cloud, 
as the source  moves from the left to the right in the HB.
In the van der Klis (2001) model, this is related to the
variation of \deltam: a higher prompt disk accretion rate is expected to push 
the inner radius towards the neutron star with a consequent larger production of seed and hotter photons for the corona, which is thus more efficiently Compton-cooled
($\kte$ decreases from regions II to IV, see Table \ref{t:regions}).
We confirm the presence of the  Fe K emission line at $\sim$ 6.7 keV,
while the edge detection (first reported by DS2000) is not 
so evident and should be treated carefully.
Indeed we find that the \eqpair\ model well fits the
data (see Table \ref{t:hybrid}) and neither Compton-reflection
nor absorption edge are  required, showing that the presence
of  such components in the spectra are 
very sensitive to the continuum model adopted.
One should be careful in claiming the presence of these components
in the case of \mbox{GX 17+2}.

\section{Conclusions}
\label{conclusions}

We have investigated two \sax\ observations of the Z source \mbox{GX 17+2} 
performed on 1997 April 3 and 21.
The persistent continuum \mbox{X-ray} emission can be described by two-component 
models, which support the well known accretion scenarios suggested by the Western and
Eastern model.
However, in the first observation, when the source was in the upper left HB of
the HID diagram, we observed a hard \mbox{X-ray} tail (up to 120 keV), 
which disappeared in the second observation  performed two weeks later, when the source moved to the 
right in the HB. Our results confirm those obtained 
by DS2000, but provide a stronger evidence for the presence of a threshold in the 
accretion rate  \deltam\  in the HB, above which the intensity of the 
hard tail  is below the instrument sensitivity or disappears.
Given that the \mbox{X-ray} luminosity is not correlated with the position of
the source in the HID, \deltam\ is not the one we ascribe to \mbox{X-ray} luminosity
but has a somewhat different meaning. We propose as working hypothesis
the model by van der Klis (2001), in which the \mbox{X-ray} luminosity is
proportional to the prompt disk accretion rate plus its own long-term
average, while motion along the Z track depends on the ratio of these
two quantities.
To testify whether this hypothesis is correct, more observational
and theoretical work is required.
The hard \mbox{X-ray} excess observed in the first observation can be described
by Comptonization  of soft seed photons ($\kts \sim 1$ keV)
off non-thermal electrons  injected with spectral index $p \sim 1.7$.
In addition, a thermal component described by a \bb\ with $\ktbb \sim 0.6$ keV
is observed. An emission line at $\sim$ 6.7 keV is found in both observations. 
The origin of the non-thermal electrons cannot be established
on the basis of the \sax\ data alone.
 We have two equally acceptable 
options: electrons from jets or electrons produced in magnetohydrodynamic turbulence 
processes occurring in the corona. Only the determination of the source 
spectrum at higher energies ($\ga 200$ keV) can provide more constraints on
the origin of the high energy excess. A step forward could be done 
by {\it INTEGRAL} through a long monitoring observation
 of the source.
Also high energy resolution observations below 10 keV, performed
 with {\it Chandra} or {\it XMM-Newton}, could be very helpful
in clarifying the origin of the Fe emission line and to
determine unambiguously the presence/absence of
an absorption edge.
On the other hand, simultaneous \mbox{X-ray}/radio observations 
 would test the possibility of a 
 correlation  between hard \mbox{X-ray} and radio emission in this source, with 
presence or not of relativistic expanding lobes.

\acknowledgements
This research is supported by the Italian Space Agency (ASI) and
Ministry of University and Scientific Research of Italy (PRIN N. 2002027145).
AAZ has been supported by KBN grants PBZ-KBN-054/P03/2001 and 1P03D01827.


\begin{thebibliography}{}
%
\bibitem[Anders \& Ebihara 1989]{Anders82}
 Anders, E., \&  Ebihara, M. 1989, Geochim. Cosmochim. Acta, 46, 2363
%
\bibitem[Asai et al. 1994]{Asai94}
 Asai, K., Dotani, T., Mitsuda, K. et al. 1994, PASJ, 46, 479
%
\bibitem[Augusteijn et al. 1992]{Aug92}
Augusteijn, T., Karatasos, K., Papadakis, M. et al. 1992, A\&A, 265, 177
%
\bibitem[Barret 2001]{Barret01}
Barret, D. 2001, Adv. Space Res., Vol. 28, p. 307
%
\bibitem[Barret \& Olive 2002]{BO02}
Barret, D., \& Olive, J. 2002, AJ, 576, 391
%
\bibitem[Boella et al.\ 1997a]{Boella97a}
Boella, G., Butler, R.C., Perola, G.C. et al. 1997a, A\&AS, 122, 299
%
\bibitem[Boella et al.\ 1997b]{Boella97b}
Boella, G., Chiappetti, L., Conti., G. et al. 1997b, A\&AS, 122, 327
%
\bibitem[Chiappetti \& Dal Fiume 1997]{CD97}
 Chiappetti, L., \& Dal Fiume, D. 1997, in Proceedings of the Fifth
 International Workshop on Data Analysis in Astronomy System, ed. V. di Ges\`u, M.J.B. Duff, A.Heck, M.C. Maccarone, L.Scarsi, \& H.U. Zimmerman, World Scientific Press, p. 101
%
\bibitem[Coppi 1999]{Coppi99}
 Coppi, P.S. 1999,  in High Energy Processes in Accreting Black Holes,
 ASP Conf. Ser. 161, ed. J. Poutanen, \& R. Svensson,  p.375
%
\bibitem[Crosby et al. 1998]{Crosby98}
Crosby, N., Vilmer, N., Lund, N., \& Sunyaev, R. 1998, A\&A, 334, 299
%
\bibitem[D'Amico et al. 2001]{Damico01}
 D'Amico, F., Heindl, W.A., Rothschild, R.E., \& Gruber, D.E. 2001, ApJ, 547, L147
%
\bibitem[Di Salvo et al. 2000]{DS2000}
  Di Salvo, T., Stella, L., Robba, N. R. et al. 2000, ApJ, 544, L119 (DS2000)
%
\bibitem[Di Salvo et al. 2001]{DS2001}
 Di Salvo, T., Robba, N.R., Iaria, R. et al. 2001, ApJ, 554, 49
%
\bibitem[Di Salvo et al. 2002]{DS2002}
 Di Salvo, T., Farinelli, R., Burderi, L. et al. 2002, A\&A, 386, 535   
%
\bibitem[Di Salvo \& Stella]{DiSalvoStella02}
 Di Salvo, T., \& Stella, L. 2002, astro-ph/0207219
%
\bibitem[Fender 2002]{Fender02}
Fender, R. 2002, in Relativistic outflows in Astrophysics, 
 ed. A.W. Guthmann, M. Georganopoulos, A. Marcowith, \& K. Manolakou, 
 Lecture Notes in Physics, vol. 589, p. 101 (astro-ph/0109502)

\bibitem[Fiore et al.\ 1999]{Fiore99}
        Fiore, F., Guainazzi, M., \& Grandi, P. 1999, Technical Report 1.2,
        {\it BeppoSAX} scientific data center, available online at 
{\tt ftp://ftp.asdc.asi.it/pub/sax/doc/
software\_docs/saxabc\_v1.2.ps}
%
\bibitem[Friedman et al. 1967]{F67}
 Friedman, H., Byram, E.T., \& Chubb, T.A. 1967, Science, 156, 374

\bibitem[Frontera et al. 1997]{Frontera97}
 Frontera, F., Costa, E., Dal Fiume, D. et al. 1997, A\&AS, 122,357
%
\bibitem[Frontera et al. 1998]{Frontera98}
 Frontera, F., Dal Fiume, D., Malaguti, G. et al. 1998, The Active X-ray Sky,
 ed. L. Scarsi, H. Bradt, P. Giommi, \& F. Fiore, Nuclear Physics. (Proc. Suppl.) 69/1--3 p. 286
%
\bibitem[Frontera et al. 2001]{Frontera01}
 Frontera, F.,  Palazzi, E., Zdziarski, A.A. et al. 2001, ApJ, 546, 1027
%
\bibitem[Hasinger et al. 1990]{H90}
 Hasinger, G., van der Klis, M., Ebisawa, K., Dotani, T., \& Mitsuda, K. 1990, A\&A, 235, 131
%
\bibitem[Hertz et al. 1990]{Hertz92}
Hertz, P., Vaughan, B., Wood, K.S. et al. 1992, ApJ, 396,201
%
\bibitem[Homan et al. 2002]{H02}
Homan, J., van der Klis, M., Jonker, P. G. et al. 2002, ApJ, 568, 878
%
\bibitem[Hoshi \& Asaoka 1993]{HA93}
 Hoshi, R., \& Asaoka, I. 1993, PASJ, 45, 567
%
\bibitem[Iaria et al. 2001]{Iaria01}
Iaria, R., Burderi, L., Di Salvo, T., La Barbera, A., \& Robba, N.R. 2001, ApJ, 547, 412
%
\bibitem[Kahn \& Grindlay 1984]{KG84}
 Kahn, S.M., \& Grindlay, J.E. 1984, ApJ, 281, 826
%
\bibitem[Kuulkers \& van der Klis]{KVDK96}
Kuulkers, E., \& van der Klis, M. 1996, A\&A, 314, 567
%
\bibitem[Magdziarz \& Zdziarski 1995]{MZ95}
 Magdziarz, P., \& Zdziarski, A.A. 1995, MNRAS, 273, 837
%
 \bibitem[Manzo et al. 1997]{Manzo97}
 Manzo, G., Giarrusso, S., Santangelo, A. et al. 1997, A\&AS, 122, 341
%
\bibitem[McClintock \& Remillard 2004]{MR04}
McClintock, J.E., \& Remillard, R.A. 2004, to appear as Chapter 4 
in Compact Stellar X-ray Sources, ed. W.H.G. Lewin, \& M.van der Klis,
 Cambridge University Press (astro-ph/0306213)
%
\bibitem[Mitsuda et al. 1984]{Mitsuda84}
Mitsuda, K, Inoue, H., Koyama, K. et al. 1984, PASJ, 36, 741
%
\bibitem[Mitsuda et al.\ 1989]{Mitsuda89}
Mitsuda, K., Inoue, H., Nakamura, N., \& Tanaka, Y.  1989, PASJ, 41, 97
%
 \bibitem[Parmar et al. 1997]{Parmar97}
 Parmar, A.N., Martin, D.D.E, Bavdaz, M. et al. 1997, A\&AS, 122, 309
%
\bibitem[Penninx et al. 1988]{Penninx88} 
 Penninx, W., Lewin, W.H.G., Zijlstra, A.A., Mitsuda, K., \& van Paradijs, J. 1988, Nature, 336, 146
%
\bibitem[Piraino et al. 2002]{Piraino02}
Piraino, S., Santanelo, A., \& Kaaret, P. 2002, ApJ, 567, 1091
%
\bibitem[Sunyaev \& Titarchuk 1980]{Sunyaev80}
         Sunyaev, R.A., \&  Titarchuk, L. 1980, A\&A, 86, 121
%
\bibitem[Sztajno et al. 1986]{Sztajno86}
Sztajno, M., van Paradijs, J., Lewin, W.H.G. 1986, MNRAS, 222, 499
%
\bibitem[Tawara et al. 1984]{Tawara84}
 Tawara, Y., Tatsumi, H., \& Tsuneo, K. 1984, PASJ, 36, 861
%
\bibitem[Titarchuk 1994]{Titarchuk94}
Titarchuk, L.G. 1994, ApJ, 434, 570
%
\bibitem[van der Klis 2001]{vdk01}
van der Klis, M. 2001, ApJ, 561, 943
%
\bibitem[van Paradijs et al. 1990]{vp90}
van Paradijs, J., Allington-Smith, J., Callanan, P. et al. 1990, A\&A, 235, 156
%
\bibitem[Vrtilek et al. 1990]{V90}
Vrtilek, S.D., Raymond, J.C., Garcia, M.R. et al. 1990, A\&A, 235, 162
%
\bibitem[Vrtilek et al. 1991]{V91a}
Vrtilek, S.D., Penninx, W., Raymond, J.C. et al., 1991a, ApJ, 376, 278
%
\bibitem[Vrtilek et al. 1991]{V91b}
 Vrtilek, S.D., McClintock, J.E., \& Seward, F.D. 1991b, ApJS, 76, 1127
%
\bibitem[White et al. 1978]{White 78} 
 White, N.E., Mason, K. O., \& Sanford, P.W. 1978, ApJ, 220, 600
%
\bibitem[White et al. 1986]{White86}
White, N. E., Peacock, A., Hasinger, G. et al. 1986, MNRAS, 218, 129 
%
\bibitem[White et al. 1988]{White88} 
White, N.E., Stella, L., \& Parmar, A.N. 1988, ApJ, 324, 363
%
\bibitem[Zdziarski et al. 1993]{Zdziarski93}
Zdziarski, A.A., Lightman, A. P., \& Macio{\l}ek-Nied\'zwiecki, A. 1993, ApJ, 414, L93
%
\bibitem[Zdziarski et al. 2001]{Zdziarski01}
Zdziarski, A.A., Grove, J.E., Poutanen, J., Rao, A.R., \&  Vadawale, S.V. 2001, ApJ, 554, L45
%
\bibitem[Zdziarski \& \gierlinski\ 2004]{ZG04}
Zdziarski, A.A., \& Gierli\'nski, M. 2004, Progr. Theor. Phys. Suppl., p. 155, in press (astro-ph /0403683)

\end{thebibliography}
\end{document}